\documentclass[12pt,prd,tightenlines,nofootinbib,showpacs,showkeys]{revtex4}
\usepackage{bm}
\usepackage{graphics}
\usepackage{rotating}
\usepackage{epsfig}
\begin{document}

\title{\begin{flushright}{\rm\normalsize SSU-HEP-06/10}\end{flushright}
FINE AND HYPERFINE STRUCTURE OF\\
$P$-LEVELS IN MUONIC HYDROGEN}
\author{A. P. Martynenko\footnote{E-mail:~mart@ssu.samara.ru}}
\affiliation{Samara State University, Pavlov Street 1, Samara 443011,
Russia}

\begin{abstract}
Corrections of orders $\alpha^5$ and $\alpha^6$ are calculated in the
fine structure interval $\Delta E^{fs}=E(2P_{3/2})-E(2P_{1/2})$
and in the hyperfine structure of the energy levels $2P_{1/2}$ and
$2P_{3/2}$ in muonic hydrogen. The obtained numerical values
$\Delta E^{fs}= 8352.08~~\mu eV$, $\Delta
\tilde E^{hfs}(2P_{1/2})=7964.36~~\mu eV$, $\Delta
\tilde E^{hfs}(2P_{3/2})=3392.59~~\mu eV$
can be considered as a reliable estimate for the comparison with
corresponding experimental data and for the extraction of the
experimental value of the Lamb shift $(2P-2S)$ in muonic hydrogen.
\end{abstract}



\maketitle

\section{Introduction}

Simple atoms play important role in the check of quantum
electrodynamics (QED), the bound state theory and precise
determination of fundamental physical constants (the fine structure
constant, the lepton and proton masses, the Rydberg constant, the
proton charge radius, etc) \cite{EGS,SGK}. On the one hand the
essential progress achieved in the last years in a more precise
determination of fundamental physical parameters \cite{MT} is
closely related with the growth of the theoretical accuracy in the
calculation of the fine and hyperfine structure of the energy levels
of hydrogenic atoms, the lepton anomalous magnetic moments (AMM) and
on the other hand with the perfection of experimental methods in the
investigation of atomic spectra of the hydrogen, muonium,
positronium \cite{Sokolov,EGS,SGK}. Light muonic atoms
\cite{GD,lma,lma1} (muonic hydrogen $(\mu p)$, muonic deuterium,
ions of muonic helium etc.) are distinguished among simple atoms
because of the strong influence of the vacuum polarization (VP)
effects \cite{U1935,S1935}, the recoil effects \cite{BG}, the
nuclear structure and polarizability effects
\cite{friar,JB,MF3,RR,M2000,P1999,M2006} on the structure of their
energy levels. So, the comparison of the theoretical value of the
Lamb shift $(2P-2S)$ in muonic hydrogen with experimental data from
the Paul Sherrer Institute (PSI) \cite{PSI,PSI1} will make it
possible to obtain a more precise value of the proton charge radius
and to check the quantum electrodynamics with the accuracy
$10^{-7}$. Considering that the energy interval
$(2^5P_{3/2}-2^3S_{1/2})$ is investigated at PSI with an accuracy
$30~ppm$, it is important to calculate precisely not only the Lamb
shift value, but also the fine and hyperfine structure of the $S$ -
wave and $P$ - wave energy levels in the atom $(\mu p)$.

The energy levels of light muonic atoms were theoretically studied
many years ago in \cite{diG,B1982,B1976,P1996,Romanov} both on the basis
of the Dirac relativistic equation and the nonrelativistic approach,
taking into account corrections by the perturbation theory (PT). In
these papers (see also references \cite{P2004,B2005}) the basic
contributions to the fine and hyperfine structure of the $P$- energy
levels in muonic hydrogen were evaluated with the accuracy
$0.001~meV$. The structure of $S$-wave states in $(\mu p)$ was
studied in \cite{M1,DB,M2,M3} accounting corrections of order
$\alpha^5$ and $\alpha^6$. In this work we continue the
investigation of the energy spectrum of $(\mu p)$ in the $P$-wave
part. The aim of the present study is to calculate such
contributions of order $\alpha^5$ and $\alpha^6$ both in the fine
and hyperfine structure of the energy states $2P_{1/2}$, $2P_{3/2}$,
which are connected with the electron vacuum polarization, the
recoil effects, the muon anomalous magnetic moment and the
relativistic corrections. The role of all these effects is crucial
to attain the desirable accuracy. Our goal also consists in the
refinement of the earlier performed calculations in
\cite{B1982,P1996} and in the derivation of the reliable numerical
estimate for the structure of $P$-wave levels in the atom  $(\mu
p)$, which can be used for the comparison with experimental data.
Modern numerical values of fundamental physical constants are taken
from Ref.\cite{MT}: the electron mass $m_e=0.510998918(44)\cdot
10^{-3}~GeV$, the muon mass $m_\mu=0.1056583692(94)~GeV$, the fine
structure constant $\alpha^{-1}=137.03599911(46)$, the proton mass
$m_p$ = 0.938272029(80)~GeV, the proton anomalous magnetic moment
$\kappa=1.792847351(28)$, the muon anomalous magnetic moment
$a_\mu=1.16591981(62)\cdot 10^{-3}$.

\section{Fine structure of $P$ - wave states}

Our approach to the investigation of the energy spectrum of muonic
hydrogen is based on the use of quasipotential method in quantum
electrodynamics \cite{MF1,MF2}, where the two-particle bound state
is described by the Schroedinger equation. The basic contribution
to the muon and proton interaction operator is determined by the Breit
Hamiltonian \cite{t4}:
\begin{equation}
H=\frac{{\bf p}^2}{2\mu}-\frac{Z\alpha}{r}-\frac{{\bf p}^4}{8m_1^3}-
\frac{{\bf p}^4}{8m_2^3}+\frac{\pi Z\alpha}{2}\left(\frac{1}{m_1^2}+
\frac{1}{m_2^2}\right)-
\end{equation}
\begin{displaymath}
-\frac{Z\alpha}{2m_1m_2r}\left({\bf p}^2+\frac{{\bf r}({\bf rp}){\bf p}}
{r^2}\right)+\Delta V^{fs}(r)+\Delta V^{hfs}(r),
\end{displaymath}
where $m_1$, $m_2$ are the muon and proton masses, $\mu=m_1m_2/(m_1+m_2)$ is
the reduced mass, $\Delta V^{fs}$ is the muon spin-orbit interaction:
\begin{equation}
\Delta V^{fs}(r)=\frac{Z\alpha}{4m_1^2r^3}\left[1+\frac{2m_1}{m_2}+2a_\mu
\left(1+\frac{m_1}{m_2}\right)\right]({\bf L}{\mathstrut\bm\sigma}_1),
\end{equation}
$\Delta V^{hfs}$ is the proton spin-orbit interaction and the interaction
of the muon and proton spins. The leading order $(Z\alpha)^4$ contribution
to the fine structure is determined by the operator $\Delta V^{fs}$.
As it follows from the Eq.(2), $\Delta V^{fs}$ includes also the recoil
effects (the Barker-Glover correction \cite{BG}) and the muon anomalous
magnetic moment $a_\mu$ correction. Let us write the fine structure interval
for the atom $(\mu p)$ in the form:
\begin{equation}
\Delta E^{fs}=E(2P_{3/2})-E(2P_{1/2})=\frac{\mu^3(Z\alpha)^4}{32 m_1^2}
\left[1+\frac{2m_1}{m_2}+2a_\mu\left(1+\frac{m_1}{m_2}\right)\right]+
\frac{5m_1(Z\alpha)^6}{256}-
\end{equation}
\begin{displaymath}
-\frac{m_1^2(Z\alpha)^6}{64m_2}+\frac{\alpha(Z\alpha)^6 \mu^3}{32\pi m_1^2}\left[\ln\frac{\mu(Z\alpha)^2}
{m_1}+\frac{1}{5}\right]+\alpha(Z\alpha)^4A_{VP}+\alpha^2(Z\alpha)^4B_{VP}.
\end{displaymath}

\begin{figure}[htbp]
\includegraphics{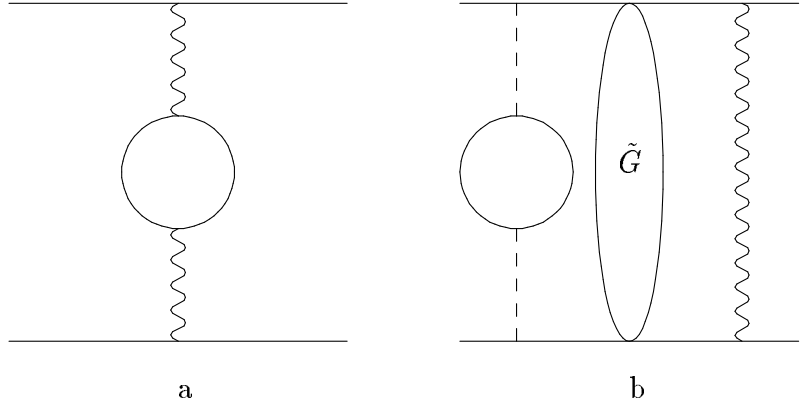}
\caption{The contribution of order $\alpha(Z\alpha)^4$ to the
fine and hyperfine structure. The dashed line corresponds the
Coulomb interaction. The wave line corresponds the fine or
hyperfine interaction. $\tilde G$ is the reduced Coulomb Green's
function.}
\end{figure}

This equation contains the relativistic correction of order
$(Z\alpha)^6$, which can be calculated on the basis of the Dirac
equation \cite{EGS,SY}, the correction of order $\alpha(Z\alpha)^6$
enhanced by the $\ln(Z\alpha)$ \cite{EY1,EY,manakov}, a number of
terms of fifth and sixth order over $\alpha$ which are determined by
the effects of the vacuum polarization \cite{U1935,S1935}. The
relativistic recoil effects of order $m_1(Z\alpha)^6/m_2$ in the
energy spectra of hydrogenic atoms were investigated in
Refs.\cite{EGS,VS1,VS2,SY,IBK}. In the fine splitting (3) they were
calculated in \cite{SY,IBK}. Additional corrections of the same
order were obtained in \cite{JP}. They don't depend on the muon
total momentum $j$ and give the contribution only to the Lamb shift.
The contributions to the coefficients $A_{VP}$ and $B_{VP}$ are
specified by the first and second order perturbation theory.
Numerical values of the terms in the expression (3) presented in the
analytical form, are depicted in Table I with the accuracy
$0.001~~\mu eV$. Such precision is related to the numerical values
of a number of contributions to the energy spectrum obtained below
from the potentials containing VP effects. The fine structure
interval (3) in the energy spectrum of electronic hydrogen is
considered for a long time as a basic test of quantum
electrodynamics \cite{SY,S1,R1972}.

The interaction operator which gives the contribution to the coefficient
$A_{VP}$, is represented by the Feynman diagrams in Fig.1.
The vacuum polarization effect leads to the modification both the Coulomb
interaction and the spin-orbit interaction in expressions (1), (2):
\begin{equation}
\Delta V^C_{VP}(r)=\frac{\alpha}{3\pi}\int_1^\infty\rho(s)ds
\left(-\frac{Z\alpha}{r}\right)e^{-2m_esr},
\end{equation}
\begin{equation}
\Delta V^{fs}_{VP}(r)=\frac{\alpha}{3\pi}\int_1^\infty\rho(s)ds
\frac{Z\alpha}{4m_1^2r^3}\left[1+\frac{m_1}{2m_2}+2a_\mu\left(1+\frac{m_1}
{m_2}\right)\right]e^{-2m_esr}(1+2m_esr)({\bf
L}{\mathstrut\bm\sigma}_1),
\end{equation}
where the spectral function $\rho(s)=\sqrt{s^2-1}(2s^2+1)/s^4$,
$m_e$ is the electron mass. The diagram (a) in Fig. 1 gives the
first order perturbation theory contribution to the fine structure
of order $\alpha(Z\alpha)^4$. Averaging the potential (2) over the
wave functions of the $2P$ - state
\begin{equation}
\psi_{2P}({\bf r})=\frac{1}{2\sqrt{6}}Wre^{-\frac{Wr}{2}}Y_{1m}(\theta,\phi),~~~
W=\mu Z\alpha,
\end{equation}
we obtain the following contribution to the interval (3):
\begin{equation}
\Delta E_1^{fs}=\frac{\mu^3(Z\alpha)^4}{32m_1^2}\left[1+\frac{m_1}{2m_2}+
2a_\mu\left(1+\frac{m_1}{m_2}\right)\right]\times
\end{equation}
\begin{displaymath}
\times\frac{\alpha}{3\pi}\int_1^\infty\rho(s)ds\int_0^\infty xdx
e^{-x\left(1+\frac{2m_es}{W}\right)}\left(1+\frac{2m_es}{W}x\right)=3.046~~\mu eV.
\end{displaymath}

Notice that the higher order corrections $\alpha^2(Z\alpha)^4$ entering
the $a_\mu$ are taken into account in this expression as well as the
recoil effects. The same order contribution $\alpha(Z\alpha)^4$
can be obtained in the second order perturbation theory (see the diagram
(b) in Fig.1). In the second order perturbation theory the energy spectrum
is determined by the reduced Coulomb Green's function whose partial
expansion has the form \cite{Palchikov}:
\begin{equation}
\tilde G_n({\bf r}, {\bf r'})=\sum_{l,m}\tilde g_{nl}(r,r')Y_{lm}({\bf n})
Y_{lm}^\ast({\bf n'}).
\end{equation}
The radial function can be represented in the form of the Sturm expansion
in the Laguerre polynomials. For the $2P$- state this function is the following:
\begin{equation}
\tilde g_{21}(r,r')=-2\mu^2\alpha zz'e^{-\frac{z+z'}{2}}\Biggl\{
\sum_{m=3}^\infty\frac{L^3_{m+1}(z)L^3_{m+1}(z')}{(m-2)(m-1)m(m+1)}-
\frac{1}{5184}\Bigl[z^4(-120+90z'-18z'^2+z'^3)+
\end{equation}
\begin{displaymath}
+z^3(3960-3270z'+756z'^2-57z'^3+z'^4)-18z^2(2160-1920z'+486z'^2-42z'^3+z'^4)+
\end{displaymath}
\begin{displaymath}
+30z(4440-4230z'+1152z'^2-109z'^3+3z'^4)\Bigr]\Biggr\},
\end{displaymath}
where $z=\mu Z\alpha r$, $L_n^m$ are the Laguerre polynomials defined by the
relation:
\begin{equation}
L_n^m(x)=\frac{e^xx^{-m}}{n!}\left(\frac{d}{dx}\right)^n\left(e^{-x}
x^{n+m}\right).
\end{equation}
To use the expression (9) in specific calculations \cite{M1,M2,M3}
we have to integrate over $z$, $z'$ and after that to sum over $m$. Another
representation for the $G_{2P}(r,r')$ was obtained in Ref.\cite{P1996} by
summing over $m$ in (9):
\begin{equation}
G_{2P}({\bf
r,r'})=-\frac{\mu^2(Z\alpha)}{36z^2z'^2}\left(\frac{3}{4\pi} {\bf
nn'}\right)e^{-\frac{z+z'}{2}}g(z,z'),
\end{equation}
\begin{equation}
g(z,z')=24z^3_{<}+36z^3_{<}z_{>}+36z^3_{<}z^2_{>}+24z^3_{>}+36z_{<}z^3_{>}+
36z^2_{<}z^3_{>}+49z^3_{<}z^3_{>}-
\end{equation}
\begin{displaymath}
-3z^4_{<}z^3_{>}-12e^z_{<}(2+z_{<}+z^2_{<})z^3_{>}-3z^3_{<}z^4_{>}+
12z^3_{<}z^3_{>}\left[-2C+Ei(z_{<}-\ln(z_{<})-\ln(z_{>})\right],
\end{displaymath}
where $z_{<}=min(z,z')$, $z_{>}=max(z,z')$, $C=0.577216...$ is the Euler
constant, $z=Wr$. Using Eqs. (11) and (12) we transform the second part
of the correction (the first part is given by Eq.(7)) of order
$\alpha(Z\alpha)^4$ to the fine structure appearing in the second order
perturbation theory as follows:
\begin{equation}
\Delta E^{fs}_2=-\frac{\alpha(Z\alpha)^4\mu^3}{3456\pi m_1m_2}\left[1+2a_\mu+
(1+a_\mu)\frac{2m_1}{m_2}\right]\times
\end{equation}
\begin{displaymath}
\times\int_1^\infty\rho(s)ds\int_0^\infty dx e^{-x\left(1+\frac{2m_es}{W}\right)}
\int_0^\infty\frac{dx'}{x'^2}e^{-x'}g(x,x')=1.928~~\mu eV.
\end{displaymath}

Let us consider the two-loop vacuum polarization contributions in
the one-photon interaction shown in Fig.2. They give the corrections
to the fine splitting of $P$- levels of order $\alpha^2(Z\alpha)^4$.

\begin{figure}[htbp]
\includegraphics{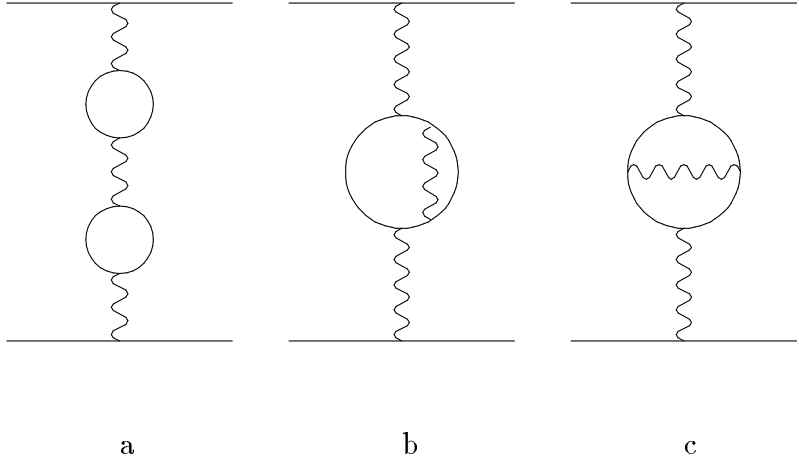}
\caption{Effects of two-loop electron vacuum polarization in the
one-photon interaction.}
\end{figure}

To find the interaction operator in the momentum representation for the
amplitude (a) in Fig.2 it is necessary to make the double change
\begin{equation}
\frac{1}{k^2}\to\frac{\alpha}{3\pi}\int_1^\infty ds\frac{\sqrt{s^2-1}(2s^2+1)}
{s^4(k^2+4m_e^2s^2)}
\end{equation}
in the photon propagator. Then in the coordinate representation the interaction
operator takes the form:
\begin{equation}
\Delta V^{fs}_{VP-VP}(r)=\frac{Z\alpha}{r^3}\left[\frac{1+2a_\mu}{4m_1^2}+
\frac{1+a_\mu}{2m_1m_2}\right]({\bf L}{\mathstrut\bm\sigma}_1)\times
\end{equation}
\begin{displaymath}
\times\left(\frac{\alpha}{3\pi}\right)^2\int_1^\infty\rho(\xi)d\xi\int_1^\infty
\rho(\eta)d\eta\frac{1}{(\xi^2-\eta^2)}\left[\xi^2(1+2m_e\xi r)e^{-2m_e\xi r}-
\eta^2(1+2m_e\eta r)e^{-2m_e\eta r}\right].
\end{displaymath}
Averaging (15) over wave functions (6), we obtain the correction to the
interval (3):
\begin{equation}
\Delta E^{fs}_3=\frac{\mu^3\alpha^2(Z\alpha)^4}{72\pi^2}\left[\frac{1+2a_\mu}
{4m_1^2}+\frac{1+a_\mu}{2m_1m_2}\right]\int_1^\infty\rho(\xi)d\xi\int_1^\infty
\rho(\eta)d\eta\frac{1}{(\xi^2-\eta^2)}\times
\end{equation}
\begin{displaymath}
\times\int_0^\infty xdx
\left[\xi^2\left(1+\frac{2m_e\xi}{W}\right)e^{-x\left(1+\frac{2m_e\xi}{W}\right)}
-\eta^2\left(1+\frac{2m_e\eta}{W}\right)e^{-x\left(1+\frac{2m_e\eta}
{W}\right)}\right]=
0.002~~\mu eV.
\end{displaymath}
The two-loop vacuum polarization amplitudes in the diagrams (b,c) of
Fig.2 can be calculated using the modification of the photon
propagator of the different type \cite{Eides}:
\begin{equation}
\frac{1}{k^2}\to \frac{2}{3}\left(\frac{\alpha}{\pi}\right)^2\int_0^1
\frac{f(v)dv}{4m_e^2+k^2(1-v^2)},
\end{equation}
\begin{equation}
f(v)=v\Biggl\{(3-v^2)(1+v^2)\left[Li_2\left(-\frac{1-v}{1+v}\right)+2Li_2
\left(\frac{1-v}{1+v}\right)+\frac{3}{2}\ln\frac{1+v}{1-v}\ln\frac{1+v}{2}-
\ln\frac{1+v}{1-v}\ln v\right]+
\end{equation}
\begin{displaymath}
\left[\frac{11}{16}(3-v^2)(1+v^2)+\frac{v^4}{4}\right]\ln\frac{1+v}{1-v}+
\left[\frac{3}{2}v(3-v^2)\ln\frac{1-v^2}{4}-2v(3-v^2)\ln v\right]+
\frac{3}{8}v(5-3v^2)\Biggr\}.
\end{displaymath}
To obtain the numerical value of the two-loop contribution in this case
it is convenient to use the coordinate representation in which the
potential corresponding to the amplitudes (b,c) in Fig.2 has the form:
\begin{equation}
\Delta V_{2-loop,VP}^{fs}(r)=\frac{2\alpha^2(Z\alpha)}{3\pi^2r^3}
\left[\frac{1+2a_\mu}{4m_1^2}+\frac{1+a_\mu}{2m_1m_2}\right]\int_0^1
\frac{f(v)dv}{1-v^2}e^{-\frac{2m_er}{\sqrt{1-v^2}}}\left(1+\frac{2m_er}
{\sqrt{1-v^2}}\right)({\bf L}{\mathstrut\bm\sigma}_1).
\end{equation}
Its contribution to the fine splitting $(2P_{3/2}-2P_{1/2})$ in muonic
hydrogen can be written in the integral form:
\begin{equation}
\Delta E^{fs}_4=\frac{\mu^3\alpha^2(Z\alpha)^4}{12\pi^2}
\left[\frac{1+2a_\mu}{4m_1^2}+\frac{1+a_\mu}{2m_1m_2}\right]\times
\end{equation}
\begin{displaymath}
\times\int_0^\infty x dx\int_0^1\frac{f(v)dv}{1-v^2}e^{-x\left(1+\frac{2m_e}
{W\sqrt{1-v^2}}\right)}\left(1+\frac{2m_e}{W\sqrt{1-v^2}}x\right)=0.021~~\mu eV.
\end{displaymath}

\begin{figure}[htbp]
\includegraphics{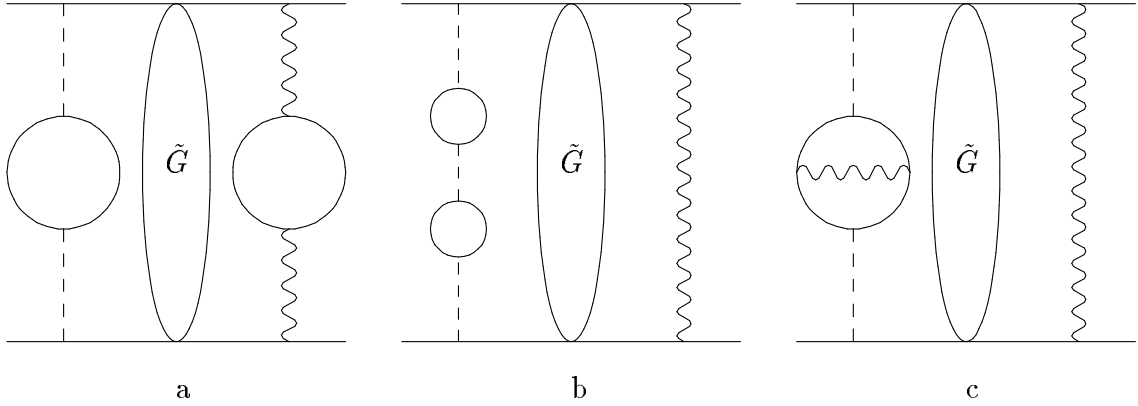}
\caption{Effects of two-loop electron vacuum polarization in the
second order perturbation theory. The dashed line corresponds to the Coulomb
interaction. The wave line corresponds to the fine or hyperfine interaction.
$\tilde G$ is the reduced Coulomb Green's function.}
\end{figure}

Two-loop vacuum polarization contributions in the second order perturbation
theory shown in Fig.3, have the same order $\alpha^2(Z\alpha)^4$.
For their calculation we can use earlier obtained relations (2), (4), (5),
(11), and the modified Coulomb potential because of the two-loop vacuum
polarization \cite{M2,M3}:
\begin{equation}
\Delta V^C_{VP-VP}(r)=\left(\frac{\alpha}{\pi}\right)^2\int_1^\infty
\rho(\xi)d\xi\int_1^\infty\rho(\eta)d\eta\left(-\frac{Z\alpha}{r}\right)
\frac{1}{\xi^2-\eta^2}\left(\xi^2e^{-2m_e\xi r}-\eta^2e^{-2m_e\eta r}\right),
\end{equation}
\begin{equation}
\Delta V^C_{2-loop,VP}=-\frac{2Z\alpha}{3r}\left(\frac{\alpha}{\pi}\right)^2
\int_0^1\frac{f(v)dv}{1-v^2}e^{-\frac{2m_er}{\sqrt{1-v^2}}}.
\end{equation}
The amplitude (a) in Fig.3 gives the following correction of order
$\alpha^2(Z\alpha)^4$ to the fine splitting:
\begin{equation}
\Delta E^{fs}_5=\frac{\mu^3\alpha^2(Z\alpha)^4}{1296\pi^2}\left[\frac{1+a_\mu}
{2m_1m_2}+\frac{1+2a_\mu}{4m_1^2}\right]\int_1^\infty\rho(\xi)d\xi
\int_1^\infty\rho(\eta)d\eta\times
\end{equation}
\begin{displaymath}
\times\int_0^\infty dx e^{-x\left(1+\frac{2m_e\xi}{W}\right)}
\int_0^\infty \frac{dx'}{x'^2}\left(1+\frac{2m_e\eta x'}{W}\right)
e^{-x'\left(1+\frac{2m_e\eta}{W}\right)}g(x,x')=0.002~~\mu eV.
\end{displaymath}
Two other contributions from the amplitudes (b), (c) in Fig.3 have
the similar integral structure. Their numerical values are included
in Table I. The summary result for the fine splitting $\Delta
E^{fs}$ in $(\mu p)$ is presented here also. It accounts the
numerous earlier performed calculations discussed in the review
article \cite{EGS} and new corrections obtained in this work.

\begin{table}[htbp]
\caption{Fine structure of $P$-wave energy levels in muonic hydrogen.}
\bigskip
\begin{ruledtabular}
\begin{tabular}{|c|c|c|}   \hline
Contribution to fine & Numerical value & Reference, \\
splitting $\Delta E^{fs}$            &of the contribution in
$\mu eV$ & equation  \\  \hline
Contribution of order $(Z\alpha)^4$ &              &         \\
$\frac{\mu^3(Z\alpha)^4}{32m_1^2}\left(1+\frac{m_1}{2m_2}\right)$ & 8329.150 &  \cite{P1996,B1982},(3) \\ \hline
Muon AMM contribution &       &           \\
$\frac{\mu^3(Z\alpha)^4}{16m_1^2}a_\mu\left(1+\frac{m_1}{m_2}\right)$ & 17.637 & \cite{P1996,B1982},(3)  \\  \hline
Contribution of order $(Z\alpha)^6$: $\frac{5m_1(Z\alpha)^6}{256}$ &0.312     &  (3),\cite{SY,IBK}  \\  \hline
Contribution of order $(Z\alpha)^6m_1/m_2$: $-\frac{m_1^2(Z\alpha)^6}{64m_2}$ &-0.028 & (3),\cite{SY,IBK}  \\  \hline
Contribution of order $\alpha(Z\alpha)^4$  &          &        \\
in the first order PT $\langle\Delta V^{fs}_{VP}\rangle$              & 3.046   & (7), \cite{P1996,B1982} \\   \hline
Contribution of order $\alpha(Z\alpha)^4$  &          &        \\
in the second order PT   & 1.928   & (13)   \\
$\langle\Delta V^'_{VP}\cdot\tilde G\cdot \Delta V^{fs}\rangle$      &     &     \\   \hline
Contribution of order $\alpha(Z\alpha)^6$  &          & \cite{EY1,manakov}       \\
$\frac{\alpha(Z\alpha)^6\mu^3}{32\pi m_1^2}\left[\ln\frac{\mu(Z\alpha)^2}{m_1}
+\frac{1}{5}\right]$ & -0.008   & \cite{EGS} \\   \hline
VP Contribution in the second &      &      \\
order PT of order $\alpha^2(Z\alpha)^4$    & 0.002   & (23)   \\
$\langle\Delta V^C_{VP}\cdot\tilde G\cdot\Delta V^{fs}_{VP}\rangle$    &   &   \\  \hline
VP Contribution from $1\gamma$ interaction &      &      \\
of order $\alpha^2(Z\alpha)^4$ $\langle\Delta V^{fs}_{VP-VP}\rangle$    & 0.002   & (16)   \\   \hline
VP Contribution from $1\gamma$ interaction &      &      \\
of order $\alpha^2(Z\alpha)^4$ $\langle\Delta V^{fs}_{2-loop,VP}\rangle$    & 0.021   & (20)   \\   \hline
VP Contribution in the second &      &      \\
order PT of order $\alpha^2(Z\alpha)^4$    & -0.002   & (21)   \\
$\langle\Delta V^C_{VP-VP}\cdot\tilde G\cdot\Delta V^{fs}\rangle$    &   &   \\  \hline
VP Contribution in the second &      &      \\
order PT of order $\alpha^2(Z\alpha)^4$    & 0.022   & (22)   \\
$\langle\Delta V^C_{2-loop,VP}\cdot\tilde G\cdot\Delta V^{fs}\rangle$    &
&   \\  \hline Summary contribution  &  8352.082   &   \\   \hline
\end{tabular}
\end{ruledtabular}
\end{table}

\section{Hyperfine structure of the energy levels $2P_{1/2}$ ¨ $2P_{3/2}$}

The leading order contribution to the hyperfine splitting of the
energy levels $2P_{1/2}$ and $2P_{3/2}$ in muonic hydrogen of order
$(Z\alpha)^4$ is determined by the following potential (the hyperfine
part of the Breit potential) \cite{t4}:
\begin{equation}
\Delta V_B^{hfs}(r)=\frac{Z\alpha(1+\kappa)}{2m_1m_2r^3}\left[1+\frac{m_1(1+
2\kappa)}{2m_2(1+\kappa)}\right]({\bf L}{\mathstrut\bm\sigma}_2)-
\frac{Z\alpha(1+\kappa)}{4m_1m_2r^3}\left[({\mathstrut\bm\sigma}_1
{\mathstrut\bm\sigma}_2)-3({\mathstrut\bm\sigma}_1{\bf n})
({\mathstrut\bm\sigma}_2{\bf n})\right],
\end{equation}
where ${\bf n}={\bf r}/r$, $\kappa$ is the proton anomalous magnetic moment.
The operator (24) does not commute with the operator of the muon total angular
momentum ${\bf J}={\bf L}+\frac{1}{2}{\mathstrut\bm\sigma}_1$. So, the energy
levels $2P_{1/2}$ and $2P_{3/2}$ are mixed and the hyperfine structure
of $P$-wave levels is more complicated.

To calculate the diagonal matrix elements $\langle 2P_{1/2}|\Delta V_B^{hfs}
|2P_{1/2}\rangle $ and $\langle 2P_{3/2}|\Delta V_B^{hfs}|2P_{3/2}\rangle $
we can use the following replacement for the operators
$({\bf s}_1{\bf s}_2)$ ¨ $({\bf L}{\bf s}_2)$ containing the nuclear spin:
\begin{equation}
{\bf s}_1\to{\bf J}\frac{\overline{({\bf s}_1{\bf J})}}{J^2},~~~
{\bf L}\to{\bf J}\frac{\overline{({\bf L}{\bf J})}}{J^2},
\end{equation}
where $\overline{({\bf s}_1{\bf J})}$, $\overline{({\bf L}{\bf J})}$
are eigenvalues of corresponding operators between the states with
equal orbital momentum $l$. Moreover, the angle averaging in the second
term of (24) can be carried out by means of the relation \cite{BS}:
\begin{equation}
\langle\delta_{ij}-3n_in_j\rangle=-\frac{1}{5}(4\delta_{ij}-3L_iL_j-3L_jL_i).
\end{equation}
The diagonal matrix elements have the following general structure:
\begin{equation}
\Delta E^{hfs}(P_{1/2})=\langle P_{1/2}|\Delta V^{hfs}|P_{1/2}\rangle=
\end{equation}
\begin{displaymath}
=E_F\left[\frac{1}{3}+\frac{a_\mu}{6}+\frac{m_1(1+2\kappa)}{2m_2(1+\kappa)}+
\frac{m_1^3}{\mu^3}A_{rel}^{1/2}(Z\alpha)^2
+A_{VP}^{1/2}\alpha+B_{VP}^{1/2}\alpha^2\right],
\end{displaymath}
\begin{equation}
\Delta E^{hfs}(P_{3/2})=\langle P_{3/2}|\Delta V^{hfs}|P_{3/2}\rangle =
\end{equation}
\begin{displaymath}
=E_F\left[\frac{2}{15}-\frac{a_\mu}{30}+\frac{m_1(1+2\kappa)}{2m_2(1+\kappa)}+
\frac{m_1^3}{\mu^3}A_{rel}^{3/2}(Z\alpha)^2
+A_{VP}^{3/2}\alpha+B_{VP}^{3/2}\alpha^2\right],
\end{displaymath}
where $E_F=(Z\alpha)^4\mu^3(1+\kappa)/3m_1m_2$ is the Fermi energy for the
level with the $n=2$. The calculation of the relativistic corrections
$A_{rel}^{1/2}$, $A_{rel}^{3/2}$ in this formalism includes the study
of two-photon, three-photon exchange diagrams and the second order
perturbation theory contributions with the Breit Hamiltonian (1), (2), (24).
More simple approach to their calculation is based on the Dirac relativistic
equation \cite{Breit,EY}. In this case the hyperfine splitting potential
has the form:
\begin{equation}
\Delta V^{hfs}_D=e{\mathstrut\bm\mu}\frac{[{\bf r}\times
{\mathstrut\bm\alpha}]}{r^3},
\end{equation}
and gives the following contributions to the hyperfine structure:
\begin{equation}
\Delta E^{hfs}_{rel}(2P_{1/2})=\frac{4(Z\alpha)(1+\kappa)}{m_2}R_{1/2},
\end{equation}
\begin{displaymath}
\Delta E^{hfs}_{rel}(2P_{3/2})=-\frac{16(Z\alpha)(1+\kappa)}{15m_2}R_{3/2},
\end{displaymath}
where the nuclear magnetic moment ${\mathstrut\bm\mu}=g_N\mu_N{\bf
s}_2$ ($\mu_N=e/2m_2$). The typical radial integrals
$R_k=\int_0^\infty g_kf_k dr$ for this case are determined by the
Dirac wave functions $f_k,g_k$ of the states $2P_{1/2}$, $2P_{3/2}$.
Accounting their exact form \cite{BS} we obtain the following values
of the relativistic corrections to the hyperfine structure of the
$P$ - wave level:
\begin{equation}
A_{rel}^{1/2}=\frac{47}{72},~~~A_{rel}^{3/2}=\frac{7}{180}.
\end{equation}
These values of the coefficients coincide with the analytical calculation of
the contribution of order $m_1^2(Z\alpha)^6/m_2$ in the hyperfine structure
of the $P$ - levels of hydrogen atom for the $n=2$, carried out in Ref.\cite{IBK}.

The fifth order contribution over $\alpha$ appears in the hyperfine splitting
as well as in the fine structure due to the electron vacuum polarization
(see diagrams (a), (b) in Fig.1). The modification of the hyperfine part
of the Breit potential from the vacuum polarization is described by the
following relation (the substitution (14) is used) \cite{P1996}:
\begin{equation}
\Delta V_{VP}^{hfs}(r)=\frac{Z\alpha(1+\kappa)}{2m_1m_2r^3}\left[1+
\frac{m_1(1+2\kappa)}{2m_2(1+\kappa)}\right]({\bf L}{\mathstrut\bm\sigma}_2)
\int_1^\infty\rho(s)ds e^{-2m_esr}\left(1+2m_esr\right)-
\end{equation}
\begin{displaymath}
-\frac{Z\alpha(1+\kappa)(1+a_\mu)}{4m_1m_2r^3}\int_1^\infty\rho(s)dse^{-2m_esr}
\Bigl[4m_e^2s^2r^2\left({\mathstrut\bm\sigma}_1{\mathstrut\bm\sigma}_2-
({\mathstrut\bm\sigma}_1{\bf n})({\mathstrut\bm\sigma}_2{\bf
n})\right)+
\end{displaymath}
\begin{displaymath}
+(1+2m_esr)\left({\mathstrut\bm\sigma}_1{\mathstrut\bm\sigma}_2-
3({\mathstrut\bm\sigma}_1{\bf n})({\mathstrut\bm\sigma}_2{\bf
n})\right) \Bigr].
\end{displaymath}
For the subsequent transformations of the diagonal matrix elements
of the operator (32) we use the relation for the angle average (26)
and the similar expression for the first term in the square brackets
of Eq.(32):
\begin{equation}
\langle\delta_{ij}-n_in_j\rangle =\frac{1}{5}[2\delta_{ij}+L_iL_j+L_jL_i].
\end{equation}
Then the contributions of the vacuum polarization in the first and second orders
PT can be written as follows:
\begin{equation}
\Delta E_1^{hfs}(P_{1/2})=E_F\frac{\alpha}{18\pi}\int_1^\infty\rho(s)ds
\int_0^\infty xdx e^{-x\left(1+\frac{2m_es}{W}\right)}\times
\end{equation}
\begin{displaymath}
\times\left[\left(1+\frac{m_1(1+2\kappa)}{2m_2(1+\kappa)}\right)\left(1+
\frac{2m_es}{W}x\right)+(1+a_\mu)\left(\frac{2m_e^2s^2x^2}{W^2}+1+\frac{2m_esx}
{W}\right)\right]=3.830~~\mu eV,
\end{displaymath}
\begin{equation}
\Delta E_1^{hfs}(P_{3/2})=E_F\frac{\alpha}{18\pi}\int_1^\infty\rho(s)ds
\int_0^\infty xdx e^{-x\left(1+\frac{2m_es}{W}\right)}\times
\end{equation}
\begin{displaymath}
\times\left[\left(1+\frac{m_1(1+2\kappa)}{2m_2(1+\kappa)}\right)\left(1+
\frac{2m_es}{W}x\right)-\frac{(1+a_\mu)}{5}\left(\frac{8m_e^2s^2x^2}{W^2}+1+\frac{2m_esx}
{W}\right)\right]=0.497~~\mu eV,
\end{displaymath}
\begin{equation}
\Delta E_2^{hfs}(P_{1/2})=E_F\frac{\alpha}{324\pi}\int_1^\infty\rho(s)ds
\int_0^\infty dx e^{-x\left(1+\frac{2m_es}{W}\right)}\times
\end{equation}
\begin{displaymath}
\times\int_0^\infty\frac{dx'}{x'^2}e^{-x'}g(x,x')
\left[2+\frac{m_1(1+2\kappa)}{2m_2(1+\kappa)}+a_\mu\right]=1.838~~\mu eV,
\end{displaymath}
\begin{equation}
\Delta E_2^{hfs}(P_{3/2})=E_F\frac{\alpha}{324\pi}\int_1^\infty\rho(s)ds
\int_0^\infty dx e^{-x\left(1+\frac{2m_es}{W}\right)}\times
\end{equation}
\begin{displaymath}
\times\int_0^\infty\frac{dx'}{x'^2}e^{-x'}g(x,x')
\left[\frac{4}{5}+\frac{m_1(1+2\kappa)}{2m_2(1+\kappa)}-\frac{a_\mu}{5}\right]=
1.278~~\mu eV.
\end{displaymath}
It should be noted that the sums of the corrections (34), (36) and
(35), (37) in the terms $A_{VP}^{1/2}\alpha$, $A_{VP}^{3/2}\alpha$
are equal $0.00025$ and $0.000056$ correspondingly. They are
slightly different from the results $0.00022$ and $0.00008$ obtained
in \cite{P1996}. Neglecting the recoil effects and muon anomalous
magnetic moment in the expressions (34) and (35) we obtain the
results which coincide with the calculation based on the Dirac
equation with the potential:
\begin{equation}
\Delta V_{VP,D}^{hfs}(r)=e{\mathstrut\bm\mu}\frac{[{\bf r}\times
{\mathstrut\bm\alpha}]}{r^3}\frac{\alpha}{3\pi}\int_1^\infty\rho(s)ds
(1+2m_esr)e^{-2m_esr}.
\end{equation}
Numerical values of the contributions $A_{VP}^{1/2}\alpha$ and
$A_{VP}^{3/2}\alpha$ in this case are equal $0.00019$ and $0.00002$
correspondingly.

\begin{table}[htbp]
\caption{Hyperfine structure of $P$-wave energy levels in muonic
hydrogen.}
\bigskip
\begin{ruledtabular}
\begin{tabular}{|c|c|c|c|}   \hline
Contribution to hyperfine& Numerical value &Numerical value &Reference, \\
splitting &of the contribution in &of the contribution in& equation  \\
    & $\Delta E^{hfs}(2P_{1/2})$,~$\mu eV$ &$\Delta E^{hfs}(2P_{3/2})$,~$\mu eV$ &     \\   \hline
Contribution of order $(Z\alpha)^4$ & 7953.195  & 3392.112    &(27),(28),\cite{P1996,B1982}\\  \hline
Muon AMM contribution   &     4.432  &    -0.886  &(27),(28),\cite{P1996,B1982}     \\   \hline
Contribution of order $(Z\alpha)^6$  & 1.092    &  0.065     & (31),\cite{IBK}   \\   \hline
Contribution of order $\alpha(Z\alpha)^4$  &3.830    &0.497      & (34),(35)       \\
in the first order PT $\langle\Delta V^{hfs}_{VP}\rangle$    &          &    &  \cite{P1996,B1982} \\   \hline
Contribution of order $\alpha(Z\alpha)^4$  &          &    &(36),(37)    \\
in the second order PT  &1.838   &0.783    &  \\
$\langle\Delta V^'_{VP}\cdot\tilde G\cdot \Delta V^{hfs}\rangle$&      &     &     \\   \hline
VP Contribution in the second &      &    & (4),(32)  \\
order PT of order $\alpha^2(Z\alpha)^4$  & 0.002 &0.002    &    \\
$\langle\Delta V^C_{VP}\cdot\tilde G\cdot\Delta V^{hfs}_{VP}\rangle$&    &   &   \\  \hline
VP Contribution of $1\gamma$ interaction & 0.003     &  0.00004   &  (39)  \\
of order $\alpha^2(Z\alpha)^4$ $\langle\Delta V^{hfs}_{VP-VP}\rangle$&    &    &   \\   \hline
VP Contribution of $1\gamma$ interaction & 0.026     &0.008   &  (40) \\
of order $\alpha^2(Z\alpha)^4$ $\langle\Delta V^{hfs}_{2-loop,VP}\rangle$&    &    &    \\   \hline
VP Contribution in the second & -0.001     &-0.0003   & (11),(21),(24)  \\
order PT of order $\alpha^2(Z\alpha)^4$    &    &  &  \\
$\langle\Delta V^C_{VP-VP}\cdot\tilde G\cdot\Delta V^{hfs}\rangle$    &   &  &  \\  \hline
VP Contribution in the second &  0.017    & 0.007   & (11),(22),(24) \\
order PT of order $\alpha^2(Z\alpha)^4$    &  &   &    \\
$\langle\Delta V^C_{2-loop,VP}\cdot\tilde G\cdot\Delta V^{hfs}\rangle$& & &
\\  \hline Summary contribution  &  7964.364   & 3392.588 &    \\   \hline
\end{tabular}
\end{ruledtabular}
\end{table}

Two-loop vacuum polarization corrections in the hyperfine part of the
potential for the $l\not =0$ obtained by means of Eqs.(14), (17)
have the integral representation:
\begin{equation}
\Delta V^{hfs}_{VP-VP}(r)=\frac{Z\alpha(1+\kappa)}{2m_1m_2r^3}\left(\frac{
\alpha}{\pi}\right)^2\int_1^\infty\rho(\xi)d\xi\int_1^\infty\rho(\eta)d\eta
\frac{1}{\xi^2-\eta^2}\times
\end{equation}
\begin{displaymath}
\times\Biggl\{\left[1+\frac{m_1(1+2\kappa)}{2m_2(1+\kappa)}\right]
({\bf L}{\mathstrut\bm\sigma}_2)\left[\xi^2(1+2m_e\xi r)e^{-2m_e\xi r}-
\eta^2(1+2m_e\eta r)e^{-2m_e\eta r}\right]-
\end{displaymath}
\begin{displaymath}
-\frac{1+a_\mu}{2}\Bigl[
\left({\mathstrut\bm\sigma}_1{\mathstrut\bm\sigma}_2-
3({\mathstrut\bm\sigma}_1{\bf n})({\mathstrut\bm\sigma}_2{\bf n})\right)
\left(\xi^2(1+2m_e\xi r)e^{-2m_e\xi r}-\eta^2(1+2m_e\eta r)
e^{-2m_e\eta r}\right)+
\end{displaymath}
\begin{displaymath}
+4m_e^2r^2\left({\mathstrut\bm\sigma}_1
{\mathstrut\bm\sigma}_2-3({\mathstrut\bm\sigma}_1{\bf n})
({\mathstrut\bm\sigma}_2{\bf n})\right)\left(\xi^4 e^{-2m_e\xi r}-
\eta^4 e^{-2m_e\eta r}\right)\Bigr]\Biggr\},
\end{displaymath}
\begin{equation}
\Delta V_{2-loop,VP}^{hfs}(r)=\frac{Z\alpha(1+\kappa)}{2m_1m_2r^3}\frac{2}
{3}\left(\frac{\alpha}{\pi}\right)^2\int_0^1\frac{f(v)dv}{1-v^2}e^{-\frac{2m_er}
{\sqrt{1-v^2}}}\times
\end{equation}
\begin{displaymath}
\times\Biggl\{\left[1+\frac{m_1(1+2\kappa)}{2m_2(1+\kappa)}\right]
\left(1+\frac{2m_er}{\sqrt{1-v^2}}\right)({\bf L}
{\mathstrut\bm\sigma}_2)-\frac{1+a_\mu}{2}\times
\end{displaymath}
\begin{displaymath}
\times\left[\frac{4m_e^2r^2}{1-v^2}
\left({\mathstrut\bm\sigma}_1{\mathstrut\bm\sigma}_2-({\mathstrut\bm\sigma}_1
{\bf n})({\mathstrut\bm\sigma}_2{\bf n})\right)+
\left(1+\frac{2m_er}{\sqrt{1-v^2}}\right)
\left({\mathstrut\bm\sigma}_1{\mathstrut\bm\sigma}_2-3({\mathstrut\bm\sigma}_1
{\bf n})({\mathstrut\bm\sigma}_2{\bf n})\right)\right]\Biggr\}.
\end{displaymath}
Omitting further details of the calculation of their expectation
values which can be performed analogously (20) and (23), we
represent in Table II numerical results of the contributions of the
potentials (39), (40). Another part of two-loop corrections of order
$\alpha^2(Z\alpha)^4$ to the hyperfine structure in the second order
PT is shown in Fig.3. We also included numerical results obtained by
means of these amplitudes in Table II.

\begin{figure}[htbp]
\includegraphics{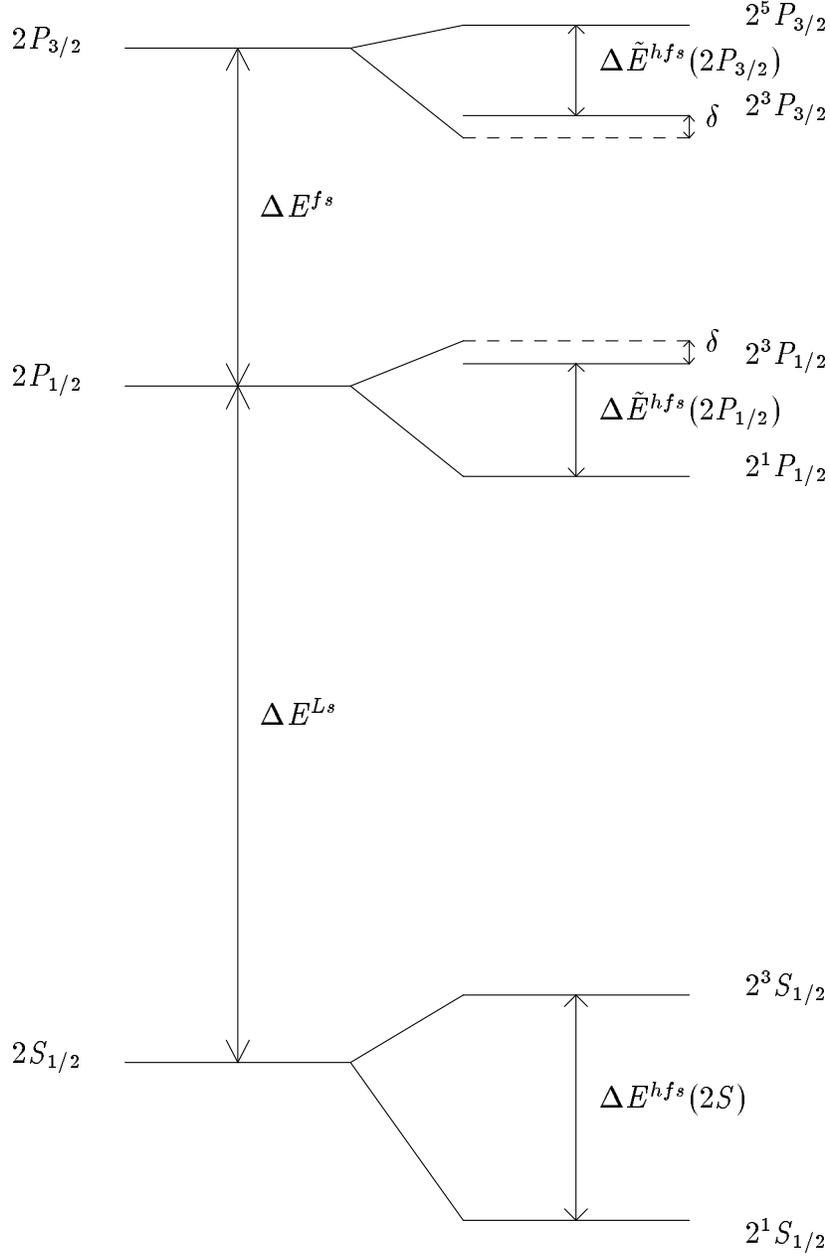}
\caption{The structure of $S$-wave and $P$-wave energy levels in muonic
hydrogen for the $n=2$.}
\end{figure}

Nondiagonal matrix element has an important role to attain the high accuracy
of the calculation of the $P$-wave levels in muonic hydrogen. We present its
general structure as follows:
\begin{equation}
\gamma=\langle^3P_{1/2}|\Delta V^{hfs}|^3P_{3/2}\rangle=E_F\left(-\frac{\sqrt{2}}{48}
\right)\left[1-a_\mu+\frac{m_1(1+2\kappa)}{m_2(1+\kappa)}+
\frac{m_1^3}{\mu^3}C_{rel}(Z\alpha)^2+C_{VP}\alpha\right],
\end{equation}
where for the simplicity we have restricted by the fifth order terms over
$\alpha$ in the vacuum polarization effects, the fifth and higher order
terms in the muon anomalous magnetic moment, the relativistic effects of order
$(Z\alpha)^6$ and the recoil effects. First three terms in the right part
of Eq. (41) appear from the potential (24). In the Dirac theory the
relativistic corrections are determined by nondiagonal radial integral:
\begin{equation}
R_{\frac{1}{2}\frac{3}{2}}=\int_0^\infty\left(g_{\frac{1}{2}}(r)f_{\frac{3}{2}}(r)
+g_{\frac{3}{2}}(r)f_{\frac{1}{2}}(r)\right)dr.
\end{equation}
After the integration in (42) which can be performed using explicit
form of the wave functions $f_{1/2,3/2}(r)$, $g_{1/2,3/2}(r)$
\cite{BS}, we obtain the coefficient $C_{rel}=9/16$. To calculate
the vacuum polarization effects we substitute the potential (32)
into $\gamma$. Then we have to calculate the matrix elements of the
following operators:
\begin{equation}
T_1=({\bf L}{\mathstrut\bm\sigma}_2),~T_2=
\left[{\mathstrut\bm\sigma}_1{\mathstrut\bm\sigma}_2-3({\mathstrut\bm\sigma}_1
{\bf n})({\mathstrut\bm\sigma}_2{\bf n})\right],~T_3=
\left[{\mathstrut\bm\sigma}_1{\mathstrut\bm\sigma}_2-({\mathstrut\bm\sigma}_1
{\bf n})({\mathstrut\bm\sigma}_2{\bf n})\right].
\end{equation}
After the angle averaging in Eq.(43) by means of Eqs.(26) and (33)
these matrix elements can be expressed in terms of the $6j$ -
symbols:
\begin{equation}
\langle T_3\rangle=-\langle T_2\rangle=-\langle T_1\rangle= =6\hat
j\hat j' \Biggl\{{{l~F~1}\atop{\frac{1}{2}~\frac{1}{2}~j}}
\Biggr\}\Biggl\{{{l~F~1}
\atop{\frac{1}{2}~\frac{1}{2}~j'}}\Biggr\}=\frac{2\sqrt{2}}{3},
\end{equation}
where the value of the total momentum $F=1$ $({\bf F}={\bf s}_2+{\bf
J})$, $l=1$, $\hat j=\sqrt{2j+1}$, $\hat j'=\sqrt{2j'+1}$, and the
numerical values of the $6j$ - symbols are taken from
\cite{Sobelman}. As a result the vacuum polarization contributions
to the nondiagonal matrix element (41) in the first and second order
PT have the form:
\begin{equation}
\gamma_1=\langle^3P_{1/2}|\Delta V^{hfs}_{VP}|^3P_{3/2}\rangle=
E_F\left(-\frac{\sqrt{2}}
{72}\right)\frac{\alpha}{\pi}\int_1^\infty\rho(s)ds\int_0^\infty xdxe^{-x\left(
1+\frac{2m_es}{W}\right)}\times
\end{equation}
\begin{displaymath}
\times\left[\left(1+\frac{m_1(1+2\kappa)}{2m_2(1+\kappa)}\right)\left(1+
\frac{2m_es}{W}x\right)-\frac{1+a_\mu}{2}\left(1+\frac{2m_es}{W}x-
\frac{4m_e^2s^2}{W^2}x^2\right)\right]=-0.617~\mu eV,
\end{displaymath}
\begin{equation}
\gamma_2=\langle^3P_{1/2}|\Delta V_{VP}^C\cdot \tilde G\cdot
\Delta V^{hfs}_{B}|^3P_{3/2}\rangle=
E_F\left(-\frac{\sqrt{2}}{2592}\right)\frac{\alpha}{\pi}
\left[1+\frac{m_1(1+2\kappa)}{m_2(1+\kappa)}-a_\mu\right]\times
\end{equation}
\begin{displaymath}
\times\int_1^\infty\rho(s)ds \int_0^\infty
xdxe^{-x\left(1+\frac{2m_es}{W}\right)}
\int_0^\infty\frac{dx'}{x'^2}e^{-x'}g(x,x')=-0.184~\mu eV.
\end{displaymath}
Putting $a_\mu=0$, $m_1/m_2=0$ in Eq.(45), we obtain the value
$-0.889$ $\mu eV$ which coincides with the calculation of this
matrix element in the Dirac theory with the potential (38).
Summary numerical value of the matrix element (41) is equal
$\gamma=-796.192~\mu eV$. It leads to the shift of the energy
levels $2^3P_{3/2}$ and $2^3P_{1/2}$ by the value $\delta=144.560~\mu eV$
as shown in Fig.4.

\section{Summary and conclusion}

In this work we calculate the QED effects in the fine and hyperfine
structure of the $2P_{1/2}$, $2P_{3/2}$ energy levels in muonic
hydrogen. The electron vacuum polarization contributions of orders
$\alpha^5$, $\alpha^6$ with the recoil corrections, the relativistic
effects of order $\alpha^6$ are considered. Numerical values of the
contributions are listed in Tables I and II. We give the references
on the papers devoted to the leading order calculation of the
structure of $P$-wave levels. For the comparison of the obtained
results with the earlier performed calculations we used the review
article \cite{EGS} containing the modern status of the
investigations in the physics of relativistic energy spectra of
simple atoms.

Let us summarize the basic particularities of the calculation performed
above.

1. Special attention in our investigation has been concentrated on the
vacuum polarization effects. For this purpose we obtain the terms of
the interaction operator in muonic hydrogen which contain the one-loop and
two-loop vacuum polarization corrections.

2. At each order over $\alpha$ we retain the recoil effects in the terms
$\sim m_1/m_2$. The experimental values of the muon and proton anomalous
magnetic moments are used \cite{MT}.

3. The calculation of the relativistic corrections to the diagonal
and nondiagonal matrix elements both for the fine and hyperfine structure
intervals is performed by means of the Dirac equation. In the second order
perturbation theory we use the compact representation for the reduced Coulomb
Green's function obtained in Ref.\cite{P1996}.

Total numerical values for the fine structure interval $\Delta E^{fs}$ (3)
and hyperfine structure splittings of $2P_{1/2}$, $2P_{3/2}$ states
are presented in Tables I,II. Accounting also our calculation of the
mixing between states $2^3P_{3/2}$ and $2^3P_{1/2}$ (the correction $\delta$),
we obtain the change of the hyperfine splittings on the $\delta=144.560~\mu eV$:
$\Delta \tilde E^{hfs}(2P_{1/2})=\Delta E^{hfs}(2P_{1/2})-\delta=
7819.804~\mu eV$, $\Delta \tilde E^{hfs}(2P_{3/2})=\Delta E^{hfs}(2P_{3/2})-\delta=
3248.028~\mu eV$. The theoretical error of the obtained results is
determined by the contributions of higher order and amounts up to $10^{-6}$.
The results of this work improve the previous calculations in
\cite{P1996,B1982,B2005} because of the investigation the effects of order
$\alpha^6$ and can be considered as a reliable estimate for the fine
and hyperfine structure intervals for the $P$- levels in muonic hydrogen.
These results are important for the experiment at $PSI$ \cite{PSI,PSI1}.
The disposition of the $P$-wave energy levels, shown in Fig.4, is determined
by the following numerical values: $-5973.27~\mu eV ~(2^1P_{1/2})$,
$1846.53~\mu eV ~(2^3P_{1/2})$, $6376.28~\mu eV ~(2^3P_{3/2})$,
$9624.30~\mu eV ~(2^5P_{3/2})$.
Accounting the Lamb shift value $(2P-2S)$ in $(\mu p)$, obtained in
Refs.\cite{B2005,P2004}, the hyperfine splitting of the $2S$- state from
\cite{M2} and the results of present study we obtain the value of the
energy interval in muonic hydrogen
$\Delta E(2^5P_{3/2}-2^3S_{1/2})=205975.6~\mu eV$.

\begin{acknowledgments}
The author is grateful to D.Bakalov, K.Pachucki and R.N.Faustov for
useful discussions. This work was supported by the Russian Fund
for Basic Research (grant No. 06-02-16821).
\end{acknowledgments}


\begin{thebibliography}{99}
\bibitem{EGS}M.I.Eides, H.Grotch, V.A.Shelyuto, Phys. Rep. {\bf 342},
62 (2001).
\bibitem{SGK}S.G.Karshenboim, Phys. Rep. {\bf 422}, 1 (2005).
\bibitem{MT}P.J.Mohr, B.N.Taylor, Rev. Mod. Phys. {\bf 72}, 351 (2000).
\bibitem{Sokolov}Yu.L.Sokolov, UFN {\bf 169}, 559 (1999).
\bibitem{GD}R.Swainson, G.W.F.Drake, Phys. Rev. A {\bf 34}, 620 (1986).
\bibitem{lma}L.A.Schaller, Z. Phys. C {\bf 56}, S48 (1992).
\bibitem{lma1}K.Jungmann, Z. Phys. C {\bf 56}, S56 (1992).
\bibitem{U1935}E.A.Uehling, Phys. Rev. {\bf 48}, 55 (1935).
\bibitem{S1935}R.Serber, Phys. Rev. {\bf 48}, 49 (1935).
\bibitem{BG}E.H.Barker, N.M.Glover, Phys. Rev. {\bf 99}, 317 (1955).
\bibitem{friar}J.L.Friar, I.Sick, Phys. Lett. B {\bf 579}, 285 (2004).
\bibitem{JB}J.Bernabeu, T.E.O.Ericson, Z.Phys. A {\bf 309}, 213 (1983).
\bibitem{MF3}R.N.Faustov, A.P.Martynenko, Phys. Rev. {\bf A67},
052506 (2003).
\bibitem{RR}R.Rosenfelder, Phys. Lett. B {\bf 463}, 317 (1999).
\bibitem{M2000}R.N.Faustov, A.P.Martynenko, Phys. Atom. Nucl. {\bf 63}, 845 (2000).
\bibitem{P1999}K.Pachucki, Phys. Rev. A {\bf 60}, 3593 (1999).
\bibitem{M2006}A.P.Martynenko, Phys. Atom. Nucl.  {\bf 69}, 1309 (2006).
\bibitem{PSI}F.Kottmann, F.Biraben, C.A.N. Conde et al., in: G.Cantatore (Ed.),
{\it Quantum Electrodynamics and Physics of the Vacuum},
QED 2000 Second Workshop Proc. New York, AIP Conf. Proc., {\bf 564}, 13
(2001).
\bibitem{PSI1}R.Pohl, A.Antognini, F.D.Amaro et al., Can. J. Phys. {\bf 83},
339 (2005).
\bibitem{diG}A.di Giacomo, Nucl. Phys. B {\bf 11}, 411 (1969).
\bibitem{B1982}E.Borie, G.A.Rinker, Rev. Mod. Phys. {\bf 54}, 67 (1982).
\bibitem{B1976}E.Borie, Z. Phys. A {\bf 278}, 127 (1976).
\bibitem{P1996}K.Pachucki, Phys. Rev. {\bf A53}, 2092 (1996).
\bibitem{Romanov}S.V.Romanov, Z.Phys. D {\bf 28}, 7 (1993).
\bibitem{P2004}A.Veitia, K.Pachucki, Phys. Rev. A {\bf 69}, 042501 (2004).
\bibitem{B2005}E.Borie, Phys. Rev. A {\bf 71}, 032508 (2005).
\bibitem{M1}R.N.Faustov, A.P.Martynenko, JETP {\bf 98}, 39 (2004).
\bibitem{DB}A.Dupays, A.Beswick, B.Lepetit et al. Phys. Rev. A {\bf 68}, 052503 (2003).
\bibitem{M2}A.P.Martynenko, Phys. Rev. A {\bf 71}, 022506 (2005).
\bibitem{M3}A.P.Martynenko, JETP {\bf 101}, 1021 (2005).
\bibitem{MF1}R.N.Faustov, A.P.Martynenko, Teor. Math. Phys. {\bf 64}, 765 (1985).
\bibitem{MF2}R.N.Faustov, A.P.Martynenko, JETP {\bf 88}, 672 (1999).
\bibitem{t4}V.B.Berestetskii, E.M.Lifshits, L.P.Pitaevskii, {\it Quantum
Electrodynamics}, Moscow, Nauka, 1980.
\bibitem{SY}J.R.Sapirstein, D.R.Yennie, in {\it Quantum Electrodynamics},
edited by T.Kinoshita, World Scientific, Singapore, p. 560, (1990).
\bibitem{EY1}G.W.Erickson, D.R.Yennie, Ann. Phys. {\bf 35}, 271 (1965).
\bibitem{manakov}N.L.Manakov, A.A.Nekipelov, A.G.Fainshtein, JETP {\bf 95},
1167 (1989).
\bibitem{EY}G.W.Erickson, D.R.Yennie, Ann. Phys. {\bf 35}, 447 (1965).
\bibitem{VS1}V.M.Shabaev, A.N.Artemyev, T.Beier, G.Soff, J. Phys. B {\bf 31},
L337 (1998).
\bibitem{VS2}A.N.Artemyev, V.M.Shabaev, V.A.Yerokhin, Phys. Rev. A {\bf 52},
1884 (1995).
\bibitem{IBK}E.A.Golosov, A.S.Yelkhovsky, A.I.Milshtein, I.B.Khriplovich,
ZhETF {\bf 107}, 393 (1995).
\bibitem{JP}U.Jentschura, K.Pachucki, Phys. Rev. A {\bf 54}, 1853 (1996).
\bibitem{S1}E.E.Salpeter, Phys. Rev. {\bf 87}, 328 (1952).
\bibitem{R1972}B.E.Lautrup, A.Peterman, E.de Rafael, Phys. Rep. {\bf 3},
193 (1972).
\bibitem{Palchikov}S.A.Zapryagaev, N.L.Manakov, V.G.Pal'chikov, Theory
of multi-charge ions with one and two electrons, Moscow, Energoatomizdat, 1985.
\bibitem{Eides}M.I.Eides, H.Grotch, V.A.Shelyuto, Phys. Rev. D {\bf 65},
013003 (2001).
\bibitem{BS}H.Bethe, E.Salpeter, Quantum mechanics of one- and two-electron
atoms, Berlin, Springer-Verlag, 1957.
\bibitem{Breit}G.Breit, Phys. Rev. {\bf 35}, 1447 (1930).
\bibitem{Sobelman}I.I.Sobel'man, Introduction to the theory of atomic
spectra, Moscow, Fizmatlit, 1963.
\end{thebibliography}
\end{document}